\documentclass[sigconf, nonacm]{acmart}

\newcommand{\ldbc}{LDBC SNB}

\usepackage{booktabs}
\usepackage{subcaption}
\usepackage{fancyvrb}

\usepackage{pifont}

\usepackage{tikz}
\usetikzlibrary{arrows.meta,positioning,fit,backgrounds,shapes.geometric}

\usepackage{booktabs}
\usepackage{makecell}
\usepackage{wrapfig}
\usepackage{paralist}

\newcommand{\myparagraph}[1]{%
  \par
  \noindent\textbf{#1}\;
}

\begin{document}
\title{Seeing the Trees for the Forest: Leveraging Tree-Shaped Substructures in Property Graphs}

\author{Daniel Aarao Reis Arturi}
\affiliation{%
  \institution{McGill University}
  \city{Montreal}
  \state{Canada}
}
\email{daniel.aaraoreisarturi@mail.mcgill.ca}

\author{Christoph Köhnen}
\affiliation{%
  \institution{University of Passau}
 \city{Passau}
  \country{Germany}
}
\email{christoph.koehnen@uni-passau.de}

\author{George Fletcher}
\affiliation{%
  \institution{Eindhoven University of Technology}
  \city{Eindhoven}
  \country{Netherlands}
}
\email{g.h.l.fletcher@tue.nl}

\author{Bettina Kemme}
\affiliation{%
  \institution{McGill University}
 \city{Montreal}
  \state{Canada}
}
\email{bettina.kemme@mcgill.ca}

\author{Stefanie Scherzinger}
\affiliation{%
  \institution{University of Passau}
  \city{Passau}
  \country{Germany}
}
\email{stefanie.scherzinger@uni-passau.de}

\begin{abstract}
Property graphs often contain tree-shaped substructures, yet they are not captured by existing proposals for graph schemas; likewise, query languages and query engines offer little-to-no native support for managing them systematically. As a first contribution, we report on a micro experiment that demonstrates the optimization potential of treating tree-shaped substructures as first class citizens in graph database systems. In particular, we show that in  systems backed by relational engines, we can achieve substantial speedups by leveraging structural indexes, as originally developed for XML databases, to accelerate path queries. Based on our findings, we put forward a vision in which tree-shaped substructures are systematically managed throughout the graph query lifecycle, from modeling and schema design to indexing and query processing, and outline arising research questions.
\end{abstract}

\maketitle

\vspace{.3cm}
\begingroup\small\noindent\raggedright
The source code, data, and experimental results have been made available at \url{https://github.com/sdbs-uni-p/trees-in-graphs}.
\endgroup

\section{Introduction}

Property graphs, as introduced in the recent edition of the ISO SQL standard, provide a convenient abstraction for representing entities and their relationships, and for expressing navigational queries over these relationships~\cite{bonifatiBook,FrancisGGLMMMPR23,GheerbrantLPR25}.
Database management systems for property graphs (GDBMSs) have become widely adopted. Some systems  are implemented as native graph engines, while others are built on top of relational database systems. 

However, planning and evaluating path queries remains a central performance challenge~\cite{MulderFY25}. Native graph systems such as Neo4j optimize graph traversals by storing neighborhood information directly alongside nodes. This allows to efficiently traverse adjacent nodes. In contrast, relational backends typically store  edges in dedicated edge tables and rely on structural joins.
Consequently, path evaluation can become very costly, especially when queries recursively traverse large subgraphs.

\emph{Motivation: there are trees in our graphs.}
An important yet underexplored observation is that many real-world property graphs actually contain a significant share of tree-shaped substructures.  Modeling data as trees is natural in many scenarios and tree-shaped substructures arise, for example, in type hierarchies, organizational structures, provenance chains, and  time-series aggregations.

\begin{figure}[tb]
  \centering

  \includegraphics[width=1\linewidth]{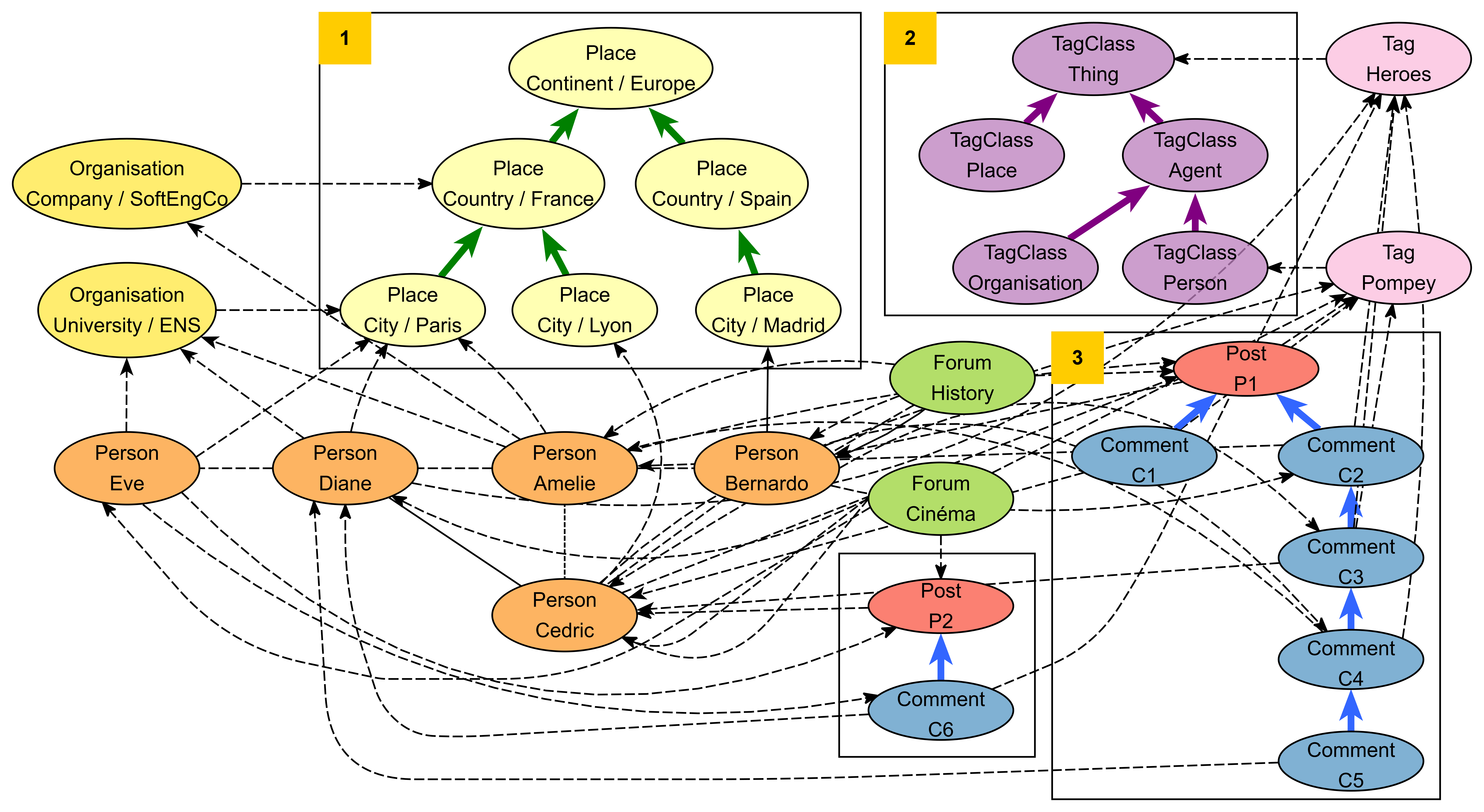}

  \caption{Excerpt\protect\footnotemark\ of \ldbc\ graph. Tree-shaped substructures framed, with thicker edges, edge labels omitted.}
  \label{fig:ldbc_excerpt}

\end{figure}

\emph{Example 1.1.}   
Consider the LDBC Social Network Benchmark (\ldbc)~\cite{10.1145/2723372.2742786}, one of the most widely used benchmarks for GDBMSs.
Figure~\ref{fig:ldbc_excerpt} shows an excerpt of this graph, adapted from the \ldbc\ documentation.
This property graph includes several tree-shaped substructures. In fact, more than 95\% of all nodes are part of some tree-shaped substructure (for \ldbc\ scale factor~1). In the figure,
frame~1 marks a hierarchy of geographical places, involving about 1.5K Place nodes.
This is an unranked, unordered tree where all edges share the same edge label and all nodes the same node label. However, the types of nodes (Continent, Country, and City), encoded as a property, differ on each level. The depth of this tree is fixed.
Frame~2 marks part of a small ontology with overall~71 TagClass nodes, for tagging posts. Frame~3 marks a tree (part of a 3M-node forest) of posts and their replies.  The statistics in Table~\ref{tab:ldbc-stats} describe the trees outlined so far, and state the share of the nodes that are part of these trees w.r.t.\ all nodes in the property graph, along with further statistics. 
Note that in this example, the edges in trees are directed ``upwards'', from child to parent. 
In the following, we treat tree-shaped substructures independently of edge orientation, requiring only that the underlying undirected structure indeed forms a tree.

\begin{figure}[t]
    \centering
    \includegraphics[width=1\linewidth]{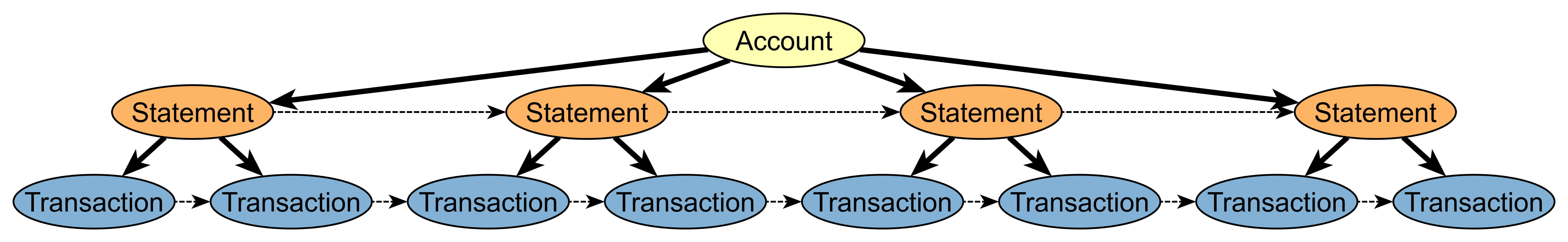}
    \caption{Time series, managing Accounts\protect\footnotemark. Edges drawn horizontally impose order among siblings. Edge labels omitted.
    }
    \label{fig:time_series}
\end{figure}

\emph{Example 1.2.} Figure~\ref{fig:time_series} shows just one further property graph in which banking transactions are grouped into statements that belong to an account. 
In this leveled tree, sibling nodes form a linked list. This imposes an order on nodes and is a common pattern for modeling time-series data.
Different from the trees discussed in Figure~\ref{fig:ldbc_excerpt}, the edges are directed downwards, from root to bottom.

Neo4j's Bloom visualization tool actually visualizes nested and sequential data in a tree or linear format. This is a strong indicator that database vendors already acknowledge tree-shaped substructures in property graphs. However, there is little-to-no system-side support for tree-shaped data.
Thus, we argue that, in contrast to the common concern of not seeing the forest for the trees, graph data management has largely overlooked the trees in the forest.

\emph{Challenge: can we take advantage of the trees in our graphs?}
The widespread use of tree-shaped substructures in property graphs raises a natural question: 
{\bf can awareness of tree-shaped substructures be leveraged to improve the usability and performance of graph queries?} 
After all, working efficiently with trees has long been a central focus in computing: interaction paradigms, query languages, and schema languages for trees can differ greatly from those for general graphs (e.g., \cite{2018Bhowmick,bonifatiBook,braga,YongBLW09}); and, fundamental algorithmic problems which are intractable on graphs often become tractable on trees (e.g., \cite{flum,GrecoLST18}). 
Indeed, the database research community has long studied efficient support for querying tree-structured data, most prominently for XML databases. Techniques for structural indexing~\cite{DGouC07}, such as hierarchical ``Dewey'' encoding or the interval-based  PrePost plane accelerate the evaluation of various query patterns, in particular the traversal of recursive paths.
These techniques exploit the structural constraints of trees in ways that are not directly applicable to general graphs.

\footnotetext{Graph adapted from \url{https://github.com/ldbc/ldbc_snb_docs}.}

\footnotetext{Graph modeled after \url{https://www.linkedin.com/pulse/modeling-longitudinaltime-seriessequential-data-neo4j-jeff-tallman/}.}

\emph{Vision: we can successfully leverage tree-structures at every stage of the graph query lifecycle.}
We present evidence that we can indeed take advantage of tree-shaped substructures in property graphs, opening the door to new research challenges at every stage of graph query processing, ranging from language and interaction design to planning and physical execution solutions.
We draw inspiration from previous successes in leveraging table-shaped substructures in databases for semantic web and NoSQL applications~\cite{DiScalaA16,PhamB16,PhamPEB15}. 

\emph{Contributions.}
Our vision paper highlights opportunities for improving the usability of graph queries for graphs with tree-shaped substructures in property graphs.
We  investigate whether structural indexing principles building on the wealth of XML knowledge can be applied to tree-shaped substructures within property graphs, and present a series of micro-experiments that demonstrates the potential of such indexes  to significantly improve query performance for GDBMS with relational engines. 
From there, we outline a wide range of research challenges.  
This starts with identifying tree-shaped substructures during schema discovery and capturing them in schema declarations. It considers handling various types of trees and supporting updates. And it requires providing the entire end-to-end query processing pipeline, including rewriting and optimization.
In general, we delineate new research lines for bridging insights from XML data management and graph database systems towards realizing substantial usability
and performance gains for a common and practically relevant class of graph workloads.

\begin{table}[tb]

\renewcommand{\arraystretch}{0.9}
\setlength{\aboverulesep}{0.3ex}
\setlength{\belowrulesep}{0.3ex}
\setlength\tabcolsep{3.7pt}
\footnotesize

\centering
\caption{Statistics on tree-shaped substructures in \ldbc\ (3.2M nodes in SF 1).
Tree size: Number of nodes.
*min/max/median over all trees; leaves not counted in fanout. **SNB/C is a subforest of the Post/Comment forest.
}
\label{tab:ldbc-stats}

\begin{tabular}{@{}lrrrrrr@{}}
\toprule
\textbf{Node label(s)}&\textbf{\#Nodes}&\textbf{Share}&\textbf{\#Trees}&\textbf{Tree size}*&\textbf{Depth}*&\textbf{Fanout}*\\
\midrule
Post/Comment&3 Mio.&96\%&1 Mio.&1/21/1&0/8/0&1/20/3\\
Comment (SNB/C)**&2 Mio.&64\%&1 Mio.&1/20/1&0/7/0&1/14/2\\
Place (SNB/P)&1,460&0.05\%&6&16/749/148&2/2/2&1/199/9\\
TagClass (SNB/T)&71&0.002\%&1&71/71/71&5/5/5&1/19/2\\
\bottomrule

\end{tabular}
\end{table}

\section{Background}
\subsection{Background}
\label{sec:background}

\subsubsection{Property-based graphs and query languages}
A \textbf{property graph}~\cite{Angles18,PG_Shema,graphQueryLanguage,bonifatiBook,FrancisGGLMMMPR23,GheerbrantLPR25} consists of \emph{nodes} connected via directional or non-directional \emph{edges} (i.e., \emph{relationships}). Nodes and edges can have one or more \emph{labels} as well as a set of \emph{properties}, each with a name and value. 
Over the years, many \textbf{graph query languages} have been developed. They have two main components (\cite{gremlin,francis2018cypher,van2016pgql,gsql,sparql}): they can contain graph and path patterns and SQL-like constructs (projections, selections, joins over graph patterns, aggregations, etc.). 
In Cypher, graph and path patterns are written in the form of \begin{small}\verb|MATCH p_var = (n_var1:N_LABEL1)-[:E_LABEL]->(n_var2)|\end{small}, where expressions in parentheses and square brackets refer to nodes resp.\ edges, and a path is visualized through the undirected ``\verb|-|'' and directed ``\verb|->|'' connectors.  
During query evaluation, qualifying nodes, edges, and paths are then bound to the variables.

\subsubsection{Indexing and query evaluation on tree-based data}
\label{sec:tree-index}

At the height of the popularity of semi-structured datastores, in particular for XML data, a plethora of indexes that capture the tree structure of the data were developed. Several of them are summarized in~\cite{DGouC07} and we refer to them as  
\emph{structural indexes}. Most well-known 
are PrePost encoding (also called interval labeling), introduced by~\cite{Dietz82} and first implemented for XML by~\cite{ZhangNDLL01}, and Dewey encoding, first implemented for XML by~\cite{TatarinovVBSSZ02}.

With PrePost, each node $a$ in a tree receives a pair of numbers $a.\text{pre}/a.\text{post}$ that correspond to the preorder and postorder traversal numbers of the node in the tree, or for XML, the positions of the opening and closing tags of the corresponding element in the XML document. With this, node~$a$ is an ancestor of node $b$ in a tree if and only if $a.\text{pre} < b.\text{pre} < a.\text{post}$. 
Table~\ref{fig:indexes} shows as example the index values for the \texttt{TagClass} tree from Figure~\ref{fig:ldbc_excerpt}, starting at the node named ``Thing''.
Looking at the \texttt{TagClass} hierarchy in Table~\ref{fig:indexes}, the TagClass node ``Agent'' is an ancestor of TagClass node ``Person'' since $\text{Agent}.\text{pre}=4<\text{Person}.\text{pre}=7<9=\text{Agent}.\text{post}$. However, $\text{Agent}.\text{pre}=4>2=\text{Place}.\text{pre}$ shows that this agent node is not an ancestor of the node with name ``Place''.

With Dewey, each node is associated with a vector (or string) of numbers to represent the path from the root to the node. If a root node has vector $1$, then its $n$ children have vectors $1.1,\dots 1.n$, and their grandchildren $1.1.1,\dots, 1.n.k_n$ (assuming that $1.n$ has $k_n$ children). 
Then, a node $a$ is an ancestor of node $b$ if and only if $a.\text{dewey}$ is a prefix of $b.\text{dewey}$.
Considering the Dewey vectors, $\text{Agent}.\text{dewey}=1.2$ is a prefix of $\text{Person}.\text{dewey}=1.2.2$, but not of $\text{Place}.\text{dewey}=1.1$. This again shows that ``Agent'' is an ancestor of ``Person'' but not of  ``Place''.

For both PrePost and Dewey, by additionally tracking the level $a.\text{lvl}$ of a node $a$ in the tree, one can easily determine that node $a$ is an ancestor of node $b$ that is $k$ hops away by adding the condition $a.\text{lvl} + k = b.\text{lvl}$.
Considering the Level column, we obtain $\text{Thing}.\text{lvl}+2=\text{Person}.\text{lvl}$,so the ancestor ``Thing'' is 2 hops away (and therefore a grandparent) of ``Person''.

\textit{Tradeoffs.}
Dewey is easier to maintain than PrePost under dynamic updates. Intuitively, when a new node (or subtree) is inserted, only the nodes in the subtrees rooted at the following sibling nodes of the new node need to update their Dewey vectors. Maintaining PrePost  is not as straightforward. Inserting a new node (or subtree) requires updating the Pre and Post values of all nodes except the nodes in the subtrees rooted at the preceding sibling nodes of the new node. 
On the other hand, PrePost  requires less storage space than Dewey vectors and provides more efficient support for checking node reachability, as number-comparisons  are cheaper than checking prefix-containment for two Dewey vectors. 

\textit{Physical query optimization.}
There is a rich literature on accelerating query evaluation with XML indexes and views for the standard XML languages XQuery and XPath~
\cite{2019Afrati,BonczGKMRT06,DGouC07,GrustJOY09,GrustKT04}. Given the significant design differences between these earlier languages and the languages proposed in the property graph model, these compilation and acceleration techniques are not immediately applicable to property graph query evaluation without further research.

\subsection{Current Indexing Techniques on Graphs}

Traditional relational indexes are widely used in processing graph queries \cite{bonifatiBook}. In addition, there exist several classes of indexes that have been specifically designed for querying graphs.

\textit{Indexing Techniques for Graph Reachability Queries.}
Reachability queries ask for the existence or exact paths that connect a source to a target node. A recent survey~\cite{ZhangBO26} presents three main indexing techniques for reachability queries. Using the tree cover approach, the potentially general graph is transformed into a DAG, then spanning trees are computed that are indexed with the techniques discussed in Section~\ref{sec:tree-index}, e.g., with PrePost encoding. A last step connects the roots of the spanning trees and considers edges that are not covered by the spanning trees. 
While these indexes can handle any kind of graph, they target reachability queries.

\textit{Path Indexing.}
Path indexes support graph queries that contain specific path patterns~\cite{FletcherPP16,KuijpersFLY21}.
Given a pattern that appears frequently in queries, a ($k$)-path index pre-computes all paths (up to length~$k$) that fulfill this specific path pattern. Should a query contain the path pattern of the index, then the relevant paths can be quickly looked up through the path index. The challenge of path indexes is the number of possible paths with a worst-case space complexity of $O(E^k)$ where $E$ is the number of edges in the graph and $k$ the maximum length of the path pattern.

\textit{Graph Indexing.}
A generalization of path indexing is a graph index that keeps track of a specific graph pattern that can be more complex than an individual path. For example, \cite{YanYH04} assumes a set of graphs and the index tracks graphs containing the graph pattern. 
This form of indexing is quite different from the approach we adopt in our upcoming study, which assumes a single graph and tags nodes to efficiently query ancestor/descendant queries.

\section{Initial Evaluation}

\label{sec:implementation}

\begin{figure}[tb]
\centering
\footnotesize

\renewcommand{\arraystretch}{0.9}
\setlength{\aboverulesep}{0.3ex}
\setlength{\belowrulesep}{0.3ex}
\setlength{\tabcolsep}{4.5pt}

\begin{subfigure}[t]{0.58\columnwidth}
\centering

\vspace{0.4ex}

\begin{tabular}{@{}r@{\hspace{4pt}}@{\hspace{4pt}}l@{\hspace{4pt}}l@{\hspace{4pt}}@{\hspace{4pt}}c@{\hspace{4pt}}@{\hspace{4pt}}r@{}}
\toprule
\textbf{node\_id} & \textbf{name} & \textbf{dewey} & \textbf{pre/post} & \textbf{lvl}\\
\midrule
1   & Thing    & 1     & 1/10 & 0 \\
304 & Place    & 1.1   & 2/3  & 1 \\
240 & Agent    & 1.2   & 4/9  & 1 \\
302 & Organis. & 1.2.1 & 5/6  & 2 \\
212 & Person   & 1.2.2 & 7/8  & 2 \\
\bottomrule
\end{tabular}

\caption{Structural indexing for the \texttt{TagClass} nodes in Figure~\ref{fig:ldbc_excerpt}, also stating Level~lvl.}
\label{fig:indexes}

\end{subfigure}
\hfill
\begin{subfigure}[t]{0.38\columnwidth}
\centering

\vspace{0.4ex}

\begin{tabular}{@{}l r c@{}}
\toprule
\textbf{Graph} & \textbf{\#Nodes} & \textbf{Min/Max} \\
               &                  & \textbf{fanout} \\
\midrule
WT 1  & 100    & 4/6   \\
WT 2  & 1,000  & 7/9   \\
WT 3  & 10,000 & 9/11  \\
DT    & 10,000 & 1/3   \\
TF    & 40     & 1/2   \\
\bottomrule
\end{tabular}

\caption{Synthetic data}
\label{table:data}
\end{subfigure}

\caption{(a) Structural index example and  (b) graph data stats}
\label{fig:tables}

\end{figure}

The goal of our evaluation is to explore the potential of PrePost and Dewey indexes to accelerate query execution on tree-shaped substructures within graphs. Given a query that traverses the tree, will an execution plan leveraging these structural indexes execute faster than the standard execution plan of the GDBMS?

\myparagraph{Prototype implementation:} We implemented both PrePost and Dewey indexes on top of three different GDBMSs: Neo4j~\cite{neo4j}, Kuzu~\cite{kuzu}, and Apache AGE~\cite{apache_age}.
Neo4j is a native GDBMS that stores graph data using specialized data structures optimized for graph traversal. Nodes and relationships are stored as fixed-size records with pointer-based linkages, allowing for efficient traversals through pointer-chasing.
Apache AGE and Kuzu have a relational execution engine with a separate table for each node and edge label. Neighborhood searches require joins across edge and node tables.

Our prototype implementations so far only support trees where all nodes and edges have the same label (such as comments and their replies in the \ldbc, but not trees like in Figure~\ref{fig:time_series}). 
For Neo4j and Kuzu, we add Pre, Post and Dewey values as ``meta''-properties to the nodes.  For Neo4j, those properties are treated just like any other properties. Kuzu automatically translates them into columns on the table that represents the nodes in the tree. For Neo4j, we build traditional indexes on these properties. As Kuzu does not support secondary indexes on attributes, we set these attributes as the primary key for these nodes to take advantage of the built-in hash index Kuzu has over primary keys.
Apache AGE is based on PostgreSQL. We add columns for Pre, Post and Dewey values to the table that contains the nodes,
and then build traditional indexes over those columns. This approach is different than for Neo4j/Kuzu because AGE stores all properties of a node in a single JSON-typed attribute, making it difficult to index a node property.

We encode Dewey values using Strings, where levels are separated by dots (e.g., ``1.2.1''). Implementing vectors as strings allows us to use a basic data type, and all GDBMS support substring search. 

\myparagraph{Queries:}
We evaluated all systems with three well-known tree-based queries:
$Q_\textit{desc}$
finds all descendants of a given node, $Q_\textit{leaf}$~finds all leaves of a given node, $Q_\textit{a\&d}$~checks whether two given nodes are in an ancestor-descendant relationship. For each GDBMS, we have three versions of these queries: A baseline Cypher query that does not use any structural index, and then a query using the PrePost resp.\ the Dewey index. For Neo4j and Kuzu, we rewrote the baseline Cypher queries to Cypher queries that use the Pre/Post/Dewey properties. For Apache AGE, we rewrote the baseline queries to equivalent SQL queries that use the index attributes we created. For example, baseline query~$Q_\textit{desc}$ to find all descendants of a given node~\verb!x! (assuming downward edges)
\begin{Verbatim}[fontsize=\small, xleftmargin=1em]
MATCH (n) - [*] -> (m) WHERE n.node_id = x RETURN m
\end{Verbatim}
is rewritten for Kuzu/Neo4j by removing the path expression and instead adding predicates on the PrePost properties to
\begin{Verbatim}[fontsize=\small, xleftmargin=1em]
MATCH (n),(m) 
WHERE n.node_id = x AND n.pre < m.pre AND m.pre < n.post 
RETURN m
\end{Verbatim}

The original and the re-written queries are available under the provided artifact link.
All queries start with one (or two) fixed nodes, and we make sure that finding these starting nodes is performed in the same way for both the baseline and the rewritten queries. Therefore, any difference in performance of these queries is due to the difference in execution on the tree component.

\myparagraph{Data:}
We use synthetically generated trees and forests (see Table ~\ref{table:data}), as well as the \ldbc\ property graph (see Table~\ref{tab:ldbc-stats}). 
In the synthetic trees, there is one common node and edge label. We show results on three shallow trees WT~1, WT~2 and WT~3 with increasing number of nodes, one large deep tree (DT), and a tiny forest~(TF) with 11 small trees that have in total 40 nodes.
Further, we use the \ldbc\ data with scale factor~1 and consider the forests of \texttt{Comment} (SNB/C), \texttt{Place} (SNB/P), resp. \texttt{TagClass} (SNB/T) nodes.
We generate three versions of each graph, for each GDBMS: one for the baseline, one for Dewey, and one for PrePost.

\myparagraph{Execution environment:}
All experiments are executed on a dual-socket server with two Intel Xeon Gold 6242R processors (40 physical cores, 80 threads, 3.1 GHz) and 192 GB RAM, running Arch Linux. Experiments run in Docker containers using Neo4j 5.26.21 (Enterprise), Kuzu 0.11.3, and Apache AGE 1.5.0 (PostgreSQL 16).

\myparagraph{Methodology:}
For Apache AGE, we measured the server-side response time and redirected the output to \texttt{/dev/null}. For Neo4j and Kuzu, end-to-end runtime was measured client-side, lacking a convenient server-side timing mechanism. Here, the results were materialized in memory but not written to standard output.

We executed each query five times and chose the median runtime, to mitigate caching effects. We set a 60-minute timeout.

\begin{figure}[tb]
    \captionsetup[subfigure]{skip=4pt}

    \centering
    \begin{subfigure}{1\linewidth}
        \centering
        \includegraphics[width=\linewidth]{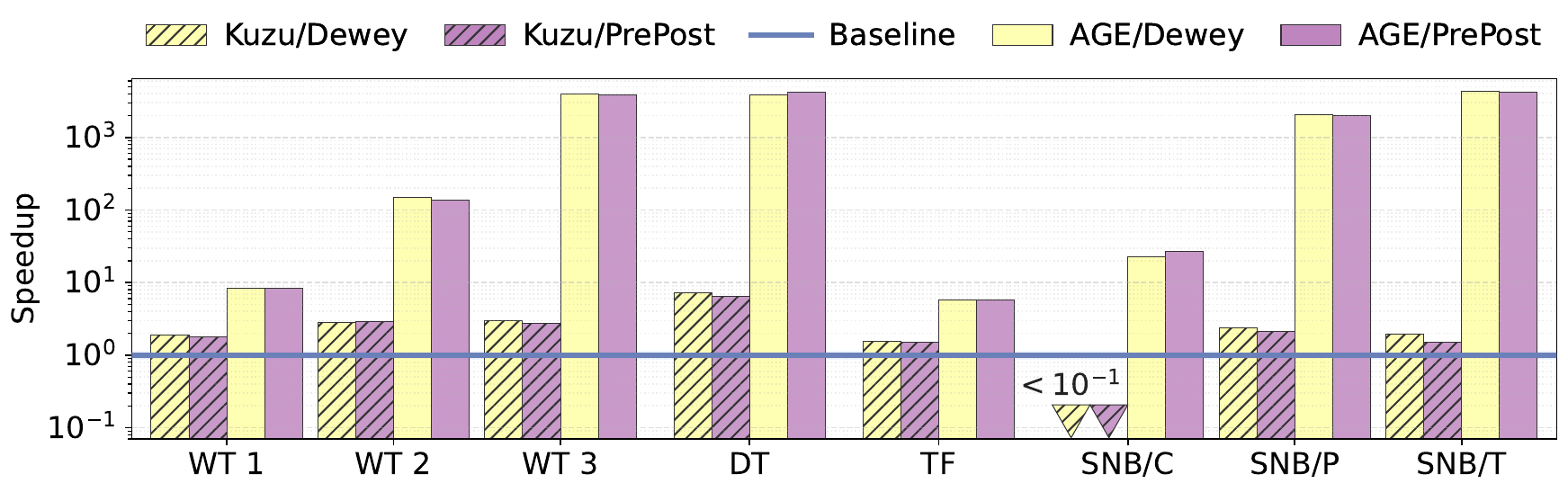}
        \caption{$Q_\textit{desc}$: Find all descendants of a given node}
    \end{subfigure}

    \vspace{0.2cm}

    \begin{subfigure}{1\linewidth}
        \centering        
        \includegraphics[width=\linewidth]{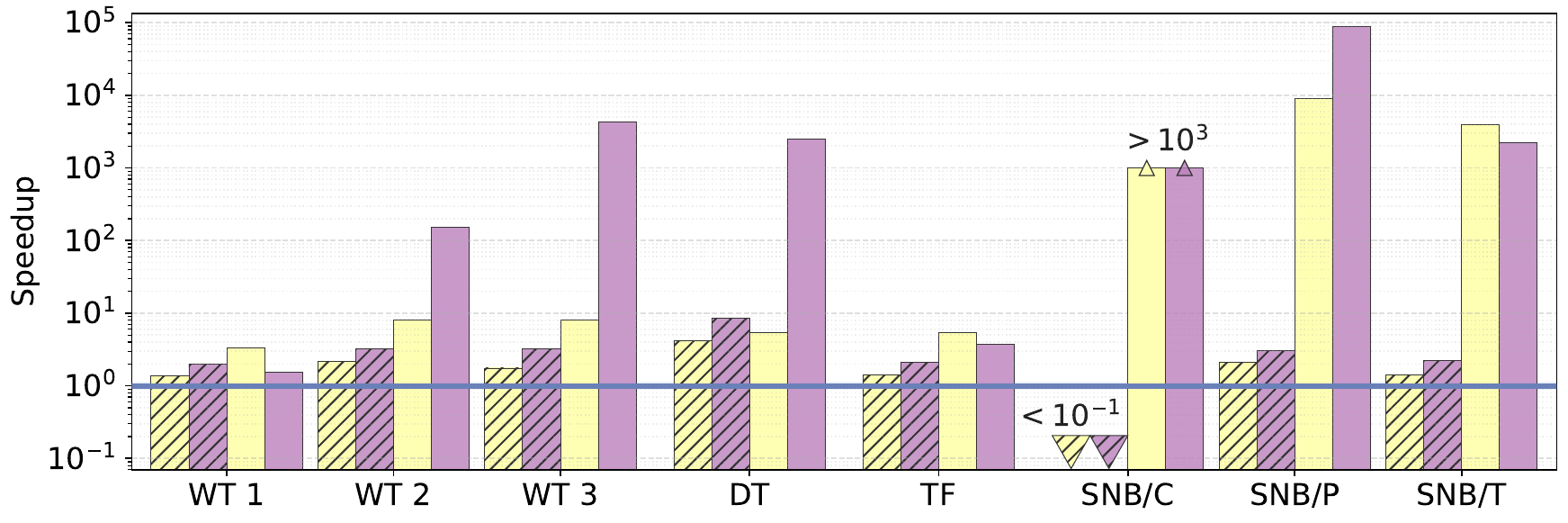}
        \caption{$Q_\textit{leaf}$: Find all leaves of a given node}
    \end{subfigure}

    \vspace{0.2cm}
    
    \begin{subfigure}{1\linewidth}
        \centering
        \includegraphics[width=\linewidth]{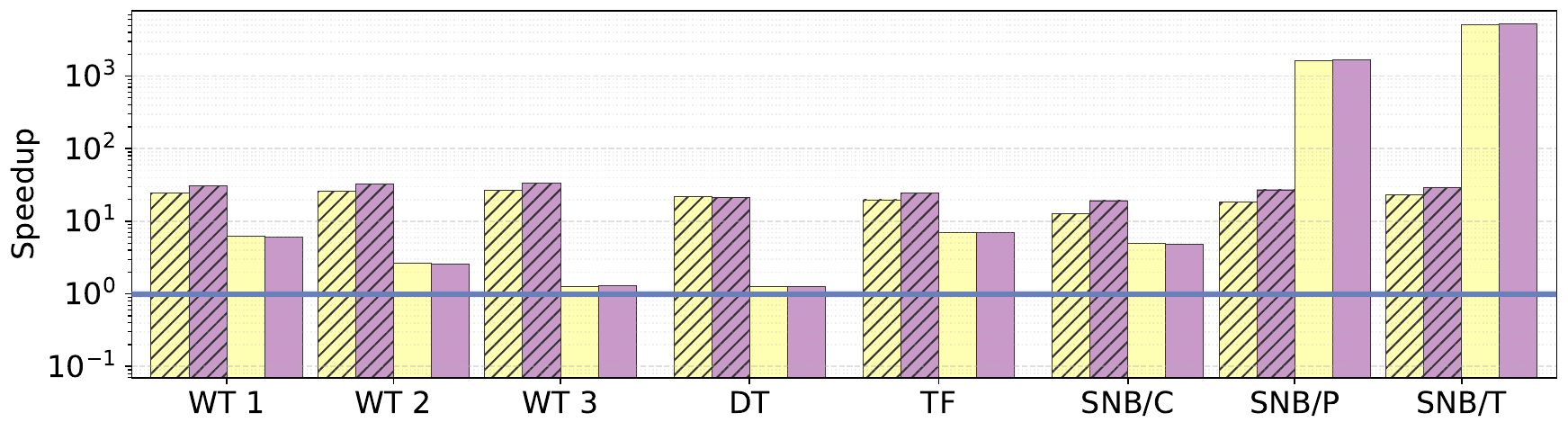}
        \caption{$Q_\textit{a\&d}$: Check whether two given nodes are ancestor\&descendant}
    \end{subfigure}
    
    \caption{Speedups in GDBMS Kuzu and Apache AGE, using Dewey and PrePost indexes on wide trees (WT~1/2/3), a deep tree (DT), a tiny forest (TF), and forests from \ldbc\ SF1.}
    \label{fig:speedups}

\end{figure}

\myparagraph{Results for Neo4j:}
In the native GDBMS Neo4j, the structural indexes largely do not improve the runtime. Runtime differences range between a 1.47x speedup and a 1.80x slowdown, depending on the query and the graph. 
Once the target node in the tree has been found, Neo4j finds adjacent nodes very quickly, as the storage layout facilitates neighborhood traversal. A proper implementation of structural indexes \emph{within} the query engine might improve performance, but we doubt it will be fundamentally better than a traditional graph traversal. Therefore, we do not further report on results for Neo4j in the following.

\myparagraph{Results for Relational Engines:}
Figure~\ref{fig:speedups} shows speedups for Kuzu and Apache AGE. Runtimes for Dewey or PrePost indexes are shown relative to the execution time of the baseline query, normalized to one and indicated by the blue horizontal line\footnote{Actual runtimes vary considerably with query, graph, index type, and GDBMS; Kuzu queries take 1-500ms,  Apache AGE queries between 2 seconds and many minutes.}.
The bars above the baseline represent an actual speedup, whereas the bars below show a slowdown. A bar reaching $10^n$ means that an indexed query gains~$n$ orders of magnitude over the baseline query. Bars labeled ``$>10^n$'' denote lower bounds on the speedup and are caused by the baseline query timing out. Triangles labeled ``$<10^{-n}$'' mark values below the visible range.

Kuzu shows a solid speedup for most queries and tree types, from~1.4 when using Dewey on the tiny forest to nearly~33 for most of the PrePost executions of query~$Q_\textit{a\&d}$. The exceptions are queries~$Q_\textit{desc}$ and~$Q_\textit{leaf}$  on  \texttt{Comment} graph SNB/C,  where a significant slowdown is observed. When comparing Dewey and PrePost for the same query and graph, they show speedups in the same order of magnitude.

In Apache AGE, the speedups are even higher for queries~$Q_\textit{desc}$ and~$Q_\textit{leaf}$ on nearly all graphs, and still significant for query~$Q_\textit{a\&d}$. There is not a single query where a slowdown was observed.  
On SNB/P and SNB/T, speedups go even above $10^3$ for all three queries. On the large synthetic trees with 10K nodes (WT~3 and DT), $Q_\textit{desc}$ and~$Q_\textit{leaf}$ also reach speedups over $10^3$ with PrePost. In most scenarios, PrePost outperforms or is at least as good as Dewey.

\myparagraph{Discussion of results:}
With the relational backends, the structural indexes show impressive performance benefits for all query types considered. Speedups tend to be higher with a larger number of nodes, although not always. The structural indexes allow the execution engine to specifically retrieve the relevant nodes without scanning the entire table. Furthermore, they avoid any kind of structural join to find neighboring nodes. For the queries considered here, there is no clear benefit of one index type over the other. We would have expected PrePost to be  more efficient than Dewey on most queries, as it executes over integers (rather than strings). However, there are several configurations where Dewey is actually faster than PrePost. 

The negative results for Kuzu on SNB/C can be partially explained by the tree structure: Most comments have no reply, so most trees consist of a single node only. The resulting small edge table leads to fast joins  in the baseline query.

Note that the actual runtimes vary greatly between Kuzu and Apache AGE. For queries~$Q_\textit{desc}$ and $Q_\textit{leaf}$, the baseline queries on the larger graphs ($\geq 10$K nodes) in Kuzu run between 500 and 5,000x faster than in Apache AGE. On WT~2, Kuzu is around 40 times faster, on the smaller graphs, the factor is below 2, but Kuzu is still faster. For query~$Q_{a\&d}$, Kuzu is around 250 times faster on \ldbc. Apache AGE only wins on the synthetic graphs for the ancestor-relationship query by a factor between~2 and~4. One reason for Kuzu's strong performance is the efficient join mechanism that is specifically designed for path queries. 
We hypothesize that the generally better performance on the baseline queries could be one of the reasons for the smaller speedup of Kuzu when applying structural indexes compared to Apache AGE.

We also conducted initial experiments with a wider range of queries and graphs. Our code repository contains for Kuzu a speedup heatmap for 10 queries  and 23 trees/forests, showing performance improvements when using structural indexes for nearly all graphs and queries. The only major exception is a query that finds all children of a given node.  Structural indexes slow down this query on some trees, but speed it up on others. Similar behavior is observed in Apache AGE. The reason is that the baseline query itself is fast: finding children requires a very small, fixed number of joins.

These initial results are very promising even with our quite simplistic prototype implementation: \textbf{Structural indexing achieves speedups of several orders of magnitude} over the baseline.

\section{Research Challenges}
\newlength{\oldintextsep}
\newlength{\oldcolumnsep}

\setlength{\oldintextsep}{\intextsep}
\setlength{\oldcolumnsep}{\columnsep}
\setlength{\intextsep}{0pt}
\setlength{\columnsep}{5pt}

We next outline research challenges spanning both systems and theory. Our vision goes well beyond transferring results from XML research, as treating tree-shaped subgraphs within property graphs as first class citizens opens up a range of genuinely new directions.

\myparagraph{How can we recognize trees in a graph?}
The structure and dependencies of the application data are typically described through a schema, and research efforts have been made to design schema languages for property graphs. Although existing proposals generally lack explicit constructs for defining hierarchical structures, they can assist in identifying tree-shaped substructures. Beyond the PG‑Schema language~\cite{PG_Shema}, a number of approaches propose visual notations in which the schema itself is represented as a graph~\cite{BonifatiDM22,Beeren2023,RabbaniLBH24}. 
Often, all nodes sharing the same set of labels are treated as belonging to a common node type and are depicted as a single schema node. Similarly, edges that share the same labels and connect nodes of the same types are abstracted into schema edges that represent edge types. Some schema languages support cardinality constraints on edge types, constraining the minimum/maximum number of outgoing/incoming edges. Previous work has proposed mechanisms to discover this schema information from an existing graph~\cite{BonifatiDM22} or to check whether a graph conforms to the schema~\cite{Beeren2023}.

The following two examples, using the schema representation from~\cite{BonifatiDM22} augmented with cardinality information,  illustrate that such schema descriptions are insufficient to define hierarchical structures, but can still provide indications for potential trees.

Recall that we distinguish the direction of edges from the hierarchies encoded in the underlying tree substructure.

\paragraph{Example 4.1.}
\begin{wrapfigure}{r}{0.25\columnwidth} 
    \includegraphics[width=0.24\columnwidth]{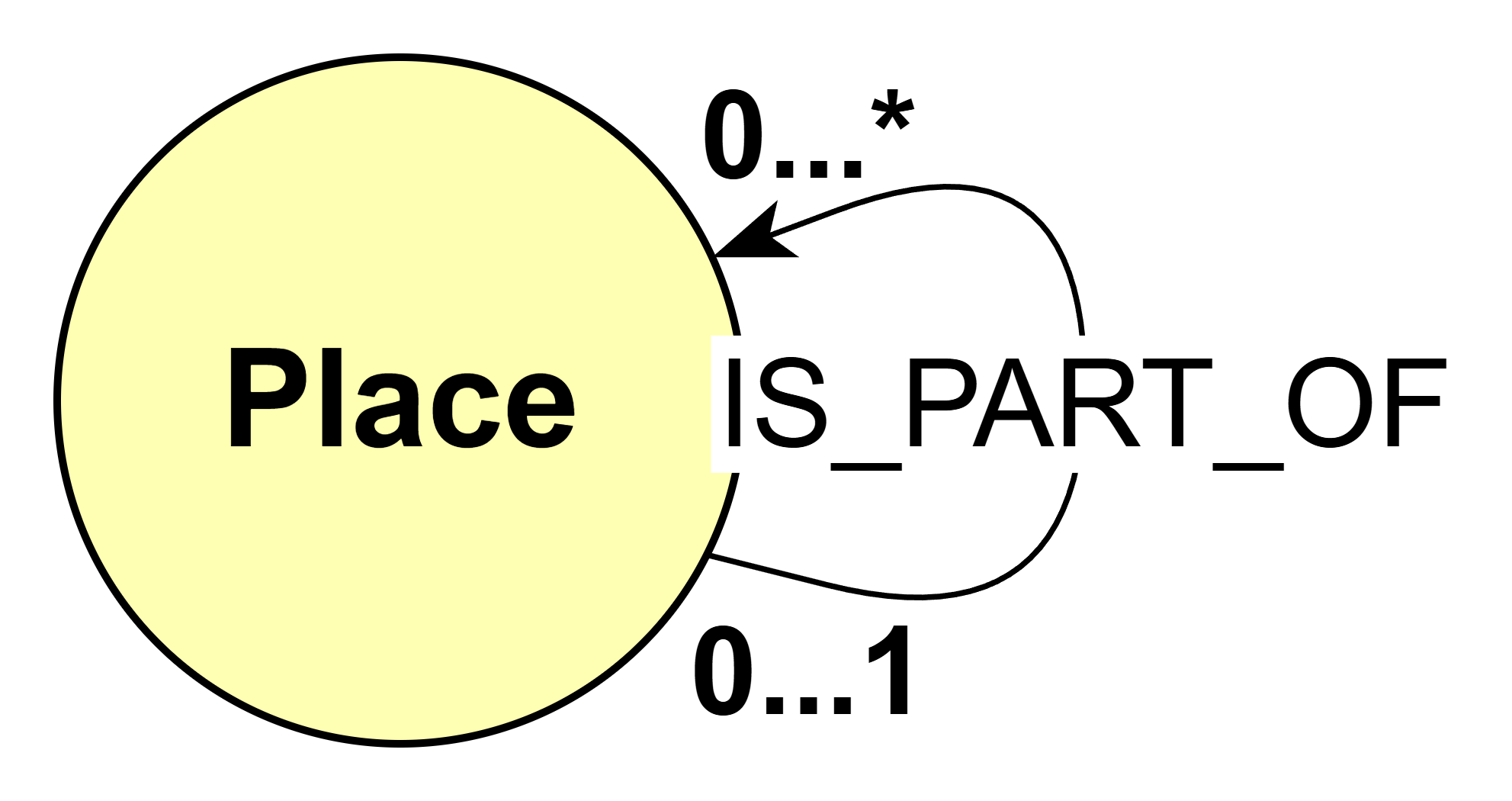}
    \caption{Place schema.}
    \label{fig:schema_place}
\end{wrapfigure}
Figure~\ref{fig:schema_place} 
shows a graph schema modeling the places in \ldbc\ (c.f.\ Figure~\ref{fig:ldbc_excerpt}).
A ``root" place (with property continent) has no parent but many children (with property country), and these country children then have city children. In fact, a tree involving only nodes and edges with the same label will lead to a schema where the node has an edge to itself with cardinalities [0...1], [0...*]. The comment node, together with the edge \texttt{REPLY\_OF} 
 of the \ldbc, will build a similar schema. However,  graphs that follow this schema might not necessarily be restricted to trees only, as the schema also allows for cycles. Nevertheless, we can use the schema as a clue and then confirm the tree structure of the graph by checking that there are no cycles. Note that the schema information does not provide any indication regarding the depth of the tree.

\paragraph{Example 4.2.}
\begin{wrapfigure}{l}{0.25\columnwidth} 
    \includegraphics[width=0.22\columnwidth]{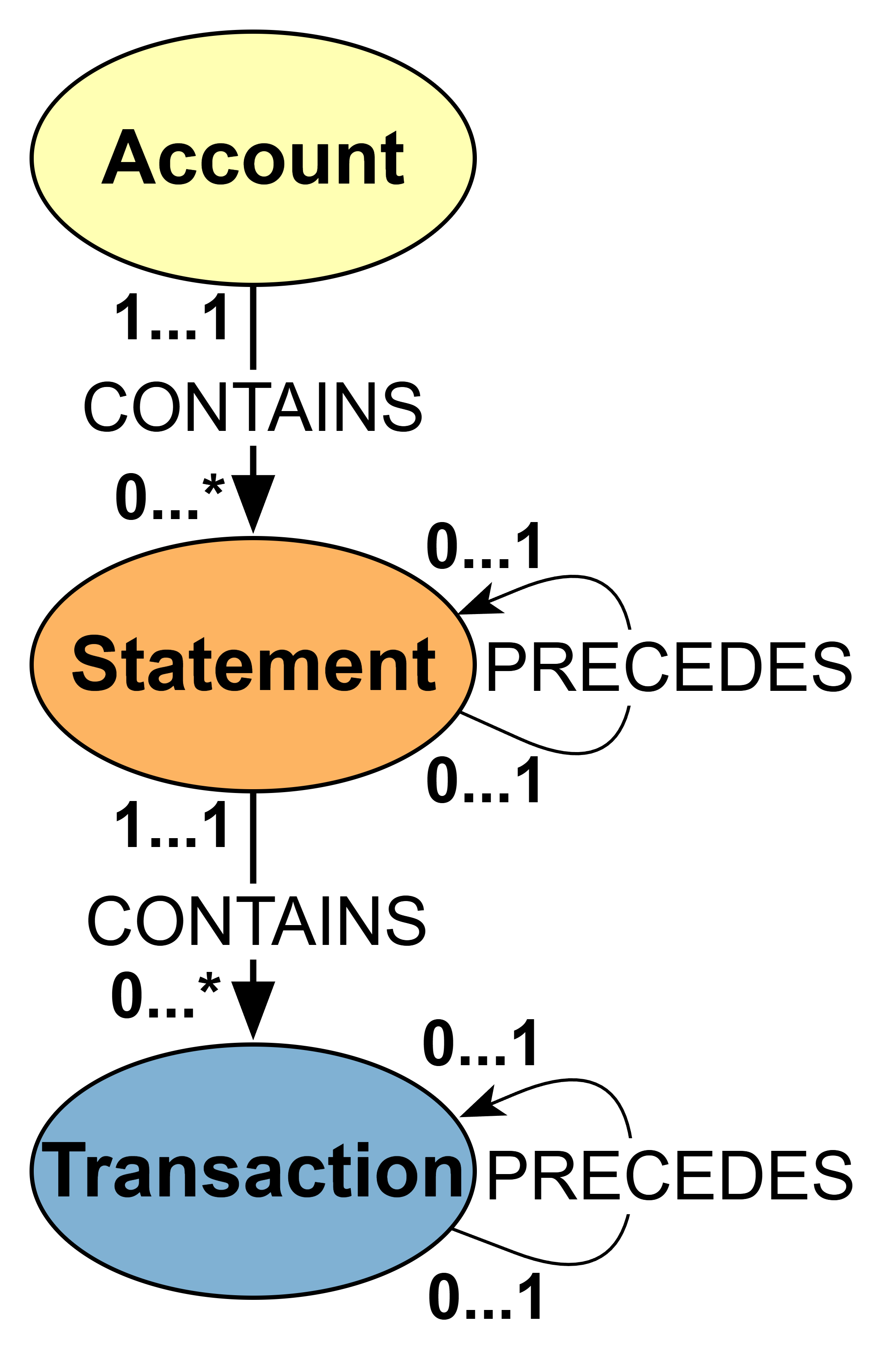}
    \caption{Time Series Schema}
    \label{fig:schema_time}
\end{wrapfigure}

Figure~\ref{fig:schema_time}
 shows the schema for the time series graph of Figure~\ref{fig:time_series}. The edges labeled~\texttt{CONTAINS} form a tree where the nodes on each level carry different labels. The height of the tree is fixed.
Here, the Statement and Transaction nodes must have a parent, as indicated by the cardinality constraint [1...1]. Compare this with~[0...1] in Figure~\ref{fig:schema_place}.
If the tree levels have different labels, the schema is sufficient to deduce the existence of trees, as cycles are impossible.

Generally, the edges in the trees can go either from parent to child or from child to parent (shown for places and for account/statement/transaction). There is no conceptual difference, and tree-based indexes should support both.

We would like to point to the self-referencing edges on Statement and Transaction with constraints [0...1], labeled \texttt{PRECEDES}. 
The cardinality constraint indicates the possibility (but not the guarantee) of a sequence (as cycles might again be possible). Sequences are another interesting sub-structure that deserves attention, as there is potential for further query optimization. In the simplest case, one can consider these sequences as trees and use tree-based indexes. However, dedicated indexes for time series databases might be even more promising.

\setlength{\intextsep}{\oldintextsep}
\setlength{\columnsep}{\oldcolumnsep}

\myparagraph{How can we specify and enforce tree schemas in a graph world?}
Ideally, the schema language allows for the explicit declaration of trees. From a syntactic point of view, we must extend textual schema languages such as PG-Schema or visual languages.
Furthermore, the tree structure must be enforced upon the insertion and deletion of edges. An insertion of an edge violates the tree structure if (a)~it would create a cycle, (b)~it would lead to a DAG, where one node has two parents. Both scenarios can be easily checked and rejected. For trees that require the nodes of certain labels to have a parent (such as Transaction and Statement in Figure~\ref{fig:time_series}), a transactional constraint must guarantee that after the insertion of a child node, a corresponding edge is added to a parent.  
The deletion of an edge might split a tree into two trees where a sub-tree now becomes an independent tree. In some cases, this would be semantically wrong, e.g., if a country is split from the corresponding continent. Ideally, a schema definition forbids this. If allowed, this would require a recalculation of the index information.

Overall, schema enforcement is not a trivial task. In fact, only few GDBMS enforce a schema at all, mostly GDBMS with a relational engine, and then, probably only because the translation to a relation schema requires some form of schema. We are not aware of any GDBMS that would enforce cardinality constraints.

\myparagraph{How can we manage trees with several node/edge types?}
Schemas can easily express trees that have different node labels and possibly also different edge labels, as we have seen in Figure~\ref{fig:time_series} and its schema in Figure~\ref{fig:schema_time}. Implementing indexes over such trees can be more complex than if node labels are homogeneous.
E.g., both Kuzu and Apache AGE store each node label in a different relational table. This breaks out prototype implementation from Section~\ref{sec:implementation}, where we add properties to a node type or columns to their tables.

For Apache AGE, one solution could be to create meta-labels such that each node in the tree has the same label. As Apache AGE creates a table for each label, all nodes with this meta-label would have a copy in this special meta-label table. This would allow us to create index attributes on this table as we did in Section~\ref{sec:implementation}. However, having nodes with two labels in Apache AGE creates redundant information (the properties of the nodes are copied over two tables). Kuzu does not allow a node to have multiple labels at all, and other GDBMS might have a completely different implementation internally. Thus, implementation will be highly system-dependent.

\myparagraph{How can we manage ordered trees?}
From an application point of view, the nodes at a certain level in the tree are often ordered. In Figure~\ref{fig:time_series} this is made explicit by having edges between siblings. Alternatively,  the nodes might have a property with increasing values among siblings, which can be common for time-series data. In fact, recent work has highlighted the links between time-series data and graphs~\cite{AmmarR0ABSKR25}. XML-inspired structural indexes have actually been designed for ordered trees and can therefore potentially achieve great speedups for queries that care about sibling order.

\myparagraph{How can we handle updates on trees?}
\label{sec:updates}
Structural indexes have well-known limitations  w.r.t.\ updates~\cite{DGouC07}  as discussed in Section~\ref{sec:background}. They hold in the same way in a GDBMS. In general, Dewey is easier to maintain. If inserts occur to the right of the most right node on a given level, no existing node needs to be updated. Such inserts might be quite common, in particular when siblings on the same level of the tree are conceptually ordered and the application mainly appends (e.g., time-series data). Yet for PrePost indexing, encodings for existing nodes need to be recalculated. If trees are large and inserts are random, both indexes might lead to a large overhead.

\myparagraph{How can we fully integrate reasoning about trees in the end-to-end query processing pipeline?}
In our prototype, we added index information as properties to nodes or as hard-coded attributes to tables, and then built traditional indexes on top. We also hard-coded queries to ensure that the index information was used in the way we intended. 
A proper implementation will face \textbf{interesting challenges}, some of which we outline below.

\subparagraph{\em Data definition languages} 
must allow for the creation of structural indexes, including those that might span node and edge types. 

\subparagraph{\em Index implementation.} The query engine needs to efficiently store indexes and link them to the original tree nodes and edges.
For tree substructures of arbitrary graphs, this will require solutions significantly different from existing approaches developed for XML.

\subparagraph{\em Query Rewriting and Optimization.} The query engine must be able to rewrite queries so that they transparently use the structural index in place. Here, query rewriting must be able to safely handle different edge directions (upwards as in Figure~\ref{fig:ldbc_excerpt} or downwards as in Figure~\ref{fig:time_series}).
Moreover, the query rewriter and the query optimizer must be aware of the index, and how an execution plan can take advantage of it. In Section~\ref{sec:implementation}, we have given some examples of queries that can take advantage of the tree structure. However, detecting that these queries can exploit a structural index is not as straightforward as it was for path queries based on XPath or XQuery where the tree-structure was clearly embedded in the queries themselves. Trees over multiple node/edge types will add additional complexity to this task. However, schema information that provides information about tree structure will likely help. In~\cite{SharmaGGL25}, the authors propose to take advantage of schema information to rewrite queries for a more efficient path execution, which can, e.g., avoid recursive queries if the depth of a path is limited and known in advance. Similar mechanisms will be needed for queries over trees.

\subparagraph{\em Query language and interaction design.} 
Property graph query languages and interaction paradigms must also be revisited to provide usable support for reasoning about trees in graphs~\cite{lics,filipov,nobre}. Application programmers are often aware of the hierarchical structure of their graph. Having specific language constructs to query those trees might not only be convenient for the programmer, but can also help with query rewriting and optimization. 

\subparagraph{\em Cost Estimation.} Current solutions are unaware of tree-shaped substructures \cite{wilco}.  To decide whether to use structural indexes, cardinality estimation techniques will need to become tree-aware.

\section{Conclusions}
Graph database systems must ultimately support arbitrary graph structures, yet we see significant untapped potential in optimizing them for tree-shaped data, which is widespread in real‑world applications. The database community has developed highly effective techniques for querying and processing hierarchical data, most notably in the context of XML. However, integrating these ideas will not be straightforward. This vision paper outlines key research opportunities, such as leveraging tree‑specific indexes, and identifies central challenges, including detecting and delineating trees within general graphs, integrating and maintaining indexes, and end‑to‑end query optimization that fully exploits tree properties.  Note that this is not comprehensive; these are just research lines that draw our interest.  We anticipate  that the community will readily identify other important lines of work.

\begin{acks}
This project/research was partly funded by the Passau International Centre for Advanced Interdisciplinary Studies (PICAIS) of the University of Passau, Germany.
\end{acks}

\clearpage

\balance

\bibliographystyle{ACM-Reference-Format}
\bibliography{references}

\end{document}